\documentclass[reprint,amsmath,amssymb,aps,prl]{revtex4-1}

\usepackage{graphicx}
\usepackage[version=3]{mhchem}

\graphicspath{{./img/}}

\begin{document}

\title{Fast exploration of chemical reaction networks}

\author{Paolo Elvati}
 \email[Corresponding author: ]{elvati@umich.edu}
  \affiliation{Department of Mechanical Engineering, University of Michigan, Ann Arbor, MI 48109-2125, USA}
\author{Angela Violi}
 \affiliation{Departments of Mechanical Engineering, Chemical Engineering,
 Biomedical Engineering, Macromolecular Science and Engineering, and Biophysics, University of Michigan, Ann Arbor, MI 48109-2125, USA}

\date{\today}

\begin{abstract}
A variety of natural phenomena comprises a huge number of competing reactions and short-lived intermediates.
Any study of such processes requires the discovery and accurate modeling of their underlying reaction network.
However, this task is challenging due to the complexity in exploring all the possible pathways and the high computational cost in accurately modeling a large number of reactions.
Fortunately, very often these processes are dominated by only a limited subset of the network's reaction pathways.
In this work we propose a novel computationally inexpensive method to identify and select the key pathways of complex reaction networks, so that high-level \textit{ab-initio} calculations can be more efficiently targeted at these critical reactions.
The method estimates the relative importance of the reaction pathways for given reactants by analyzing the accelerated evolution of hundreds of replicas of the system and detecting products formation.
This acceleration-detection method is able to tremendously speed up the reactivity of uni- and bimolecular reactions, without requiring any previous knowledge of products or transition states.
Importantly, the method is efficiently iterative, as it can be straightforwardly applied for the most frequently observed products, therefore providing an efficient algorithm to identify the key reactions of extended chemical networks.
We verified the validity of our approach on three different systems, including the reactivity of \textit{t}-decalin with a methyl radical, and in all cases the expected behavior was recovered within statistical error.

\end{abstract}


\maketitle

Many industrial and natural processes are the result of thousands of interdependent reactions.
Seemingly, a model of these processes demands all these reactions to be taken into account without regard to their ultimate importance.
Despite the fast growth of computational resources of the last few decades, this extensive modeling remains a daunting task except when massive simplifications are made.
This situation mostly originates from two problems: high computational cost of the methods that need to be employed to accurately model each reaction \cite{young_computational_2004}, and from the difficulty of finding and systematically exploring all the possible reactive pathways.

Luckily, in many cases not all the possible reactions have the same importance,
as often many reaction pathways contribute only marginally to the products'
formation.
This fact has been leveraged to build a simplified model, often called reduced mechanism \textsc{rm}, which captures the key aspects of a more detailed description while making it more tractable \cite{okino_simplification_1998}.
A \textsc{rm} is built on two distinct components, a subset of reactions and their corresponding rates; it is generally constructed through a top-down approach, where the complete reaction network is reduced to a more manageable subset \cite{tomlin_mathematical_1997}.
However except for the most simple cases, this approach involves important simplifications in the calculations of the rates \cite{susnow_rate-based_1997},
which contrasts with the high-level accuracy required to correctly model
chemical reactions \cite{golden_pressure-_2011,pu_multidimensional_2006}.
The high sensitivity of the \textsc{rm} to the methods used to compute the rates $k$ stems from the exponential dependency from the free energy \textsc{fe} difference between products and transition state $\Delta G^{\ddagger}$ \cite{truhlar_current_1996}:
\begin{equation}
	\label{E.TST}
	k = \kappa \frac{k_BT}{h} \exp(-\beta \Delta G^{\ddagger}),
\end{equation}
where $\kappa$ is the transmission coefficient, $k_B$ is the Boltzmann
constant, $T$ is the temperature, $h$ is the Planck's constant and
$\beta=(k_BT)^{-1}$.
From Eq. \ref{E.TST} follows that any error in computing $\Delta G^{\ddagger}$ is likely to have important effects on the values of the derived rates, which may have catastrophic repercussions if the error affects an early branching of a reaction network.
For this reason, instead of using one of the several existing methods developed
to compute the reaction dynamics  \cite{dellago_transition_1998,
wang_efficient_2001, erp_novel_2003, barducci_well-tempered_2008,
tiwary_metadynamics_2013} on all the possible reactions, we developed a new technique to identify and select the most frequent pathways of a reaction network with minimal computational effort by using an acceleration-detection scheme.
The rates of these primary pathways can then be computed with accurate \emph{ab-initio} techniques in order to build the \textsc{rm}.

To develop our approach, we started from the observation that for any given reactant(s), all the reaction pathways and rates can be recovered by simply observing the behavior of a large number of replicas of the same system for a long time and counting the occurrences of each reaction.
However, this method is not practical because it requires several hundred long simulations for each reactant, due to the high energy barriers commonly involved in chemical reactions.
To make this idea applicable we employed an acceleration-detection scheme, where the dynamics of all the system's replicas are accelerated until a reaction is detected. 
The simulations are then interrupted and the frequency of each reactive pathway is calculated.
This method has the advantage that it does not require \emph{a priori} knowledge of the pathways or transition states, and does not rely on the life time or stability of the products, since the simulations are interrupted as soon as the reaction happens, and therefore this approach can also handle pathways where chemical activation plays an important role.
By repeating the same procedure only for the most frequently observed reactions, the key pathways of the entire reaction network are obtained without the need to either map the complete reaction network or arbitrarily select pathways.
Although this approach can be applied to different kind of systems, in the following we will show its implementation for reactive systems in gas phase.

Since the idea of accelerating systems to overcome energetic barriers is not
new \cite{chipot_free_2007}, we used Metadynamics \cite{laio_escaping_2002}, an
already well-tested method \cite{barducci_metadynamics_2011} that uses a
history-dependent bias to favor the exploration of new states.
While Metadynamics was originally introduced to reconstruct \textsc{fe}
landscapes, here we employ it only to accelerate the system dynamics,
simplifying the requirements on the definition of the collective variables.
In particular, we use the algorithm to bias the potential energy, forcing the
reacting molecule to experience energy fluctuations typical of higher
temperatures \cite{bonomi_enhanced_2010}.
This approach can be viewed as a way to force the molecule under investigation
to experience several collisions with virtual particles that only increase its
internal energy, effectively accelerating its reactions in the high-pressure
regime.

With the addition of the potential energy bias $\varPhi$, the effective \textsc{fe} experienced by the system becomes $\Delta G_j^{\ddagger} + \varPhi$, which substituted in Eq. \ref{E.TST} can be used to compute the probability $p_i$ to observe a specific reaction $i$:
\begin{equation}
\label{E.Prob}
p_i = \frac {\kappa_i M_i \exp(-\beta \Delta G_i^{\ddagger})}
						{\sum_j \kappa_j M_j  \exp(-\beta \Delta G_j^{\ddagger})}
\text{,}
\end{equation}
where $M_i$ is the pathway degeneracy of the $i$-th reaction, and the summation is performed over all the possible reactions.

While biasing the potential energy is an effective and general way to accelerate the system reactivity \cite{michel_tracing_2009}, at the same time it is not an appropriate quantity to use to monitor the evolution of the reactions, since in many cases it is unable to distinguish between products and reactants.
A more apt choice for the reaction detection is the measure of the molecular connectivity, like the recently introduced SPRINT (Social PeRmutation INvarianT) coordinates \cite{pietrucci_graph_2011}.
This class of reaction coordinates defines the connectivity of a system of N atoms with a N-dimensional vector by using spectral graph theory and including both local and long-range system topology information.
SPRINT coordinates have already been successfully used to differentiate and cluster molecular structures \cite{lai_stochastic_2014}, by considering the evolution of the Euclidean norm of the difference between instantaneous and average value of the SPRINT coordinates.
Moreover because each reaction is interrupted as soon as a clear change in the connectivity is observed, there is no requirement for a careful selection of the SPRINT parameters in order to have a weak connectivity even after a bond is broken.

\begin{figure}
\includegraphics[width=3.4in]{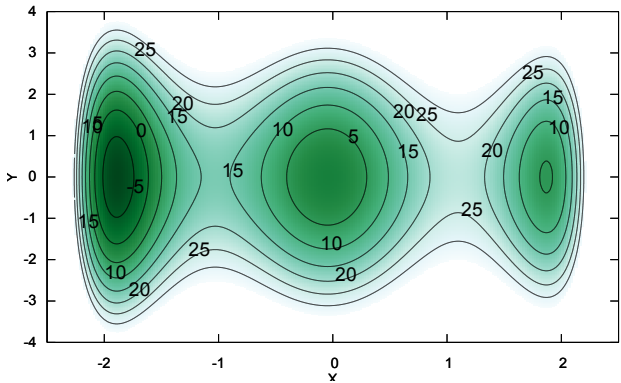}
\caption{Example of the two dimensional potential experienced by the single atom ($a_x=3,b_x=21,c_x=36,d_x=3.3,a_y=3$).
The energy is expressed in kcal/mol and the contour lines are separated by 5 energy units.
	\label{F.2dpot}}
\end{figure}

To illustrate the salient details of our acceleration-detection approach,
we present three different applications of our method.
We initially tested our technique on a simple, two-dimensional system by studying the evolution of a single atom under the effect of an analytical potential shaped to reproduce three minima separated by two asymmetric barriers, as shown in Fig. \ref{F.2dpot}.
The simulations were started from the central basin and the rates of the formation of ``products'' were computed by monitoring the position of the
particles on the x axis, for in this simple case there is no chemical connectivity that justifies the use of SPRINT coordinates.
By tweaking the parameters that define the potential, we created several scenarios with different barrier heights (from about 4 to 80 times $k_BT$) and different degrees of asymmetry between the negative and positive basins (with a difference between the two barriers ranging from 0 to about 15 $k_BT$).
We analyzed the effect of the thermostat and a variety of Metadynamics parameters, in particular deposition rate, bias shape and bias factor, when employing the well-tempered version \cite{barducci_well-tempered_2008}.
\begin{figure}
\includegraphics[width=3in]{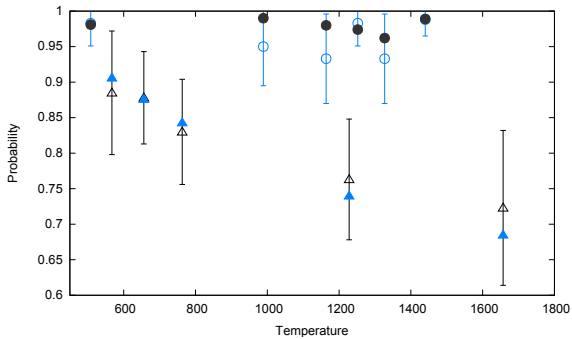}
\caption{Comparison between theoretical (full symbols) and computed (empty symbols) probability for the transition from the central ($-1 < x < 1$) to the negative basin ($x < -1$) for different potentials.
Data from simulations performed with (triangles) and without a thermostat (circles); vertical bars show the 95\% confidence interval.
	\label{F.2dres}}
\end{figure}
In all cases we found an excellent agreement between the pathway probability computed with our approach and the one predicted by using the analytical values of $\Delta G^{\ddagger}$ in Eq. \ref{E.Prob}.
A subset of the results (selected for clarity) is shown in Fig. \ref{F.2dres});
additional details can be found in the supplementary material.

As a second example we considered the first step of unimolecular decomposition of ethane at high temperatures.
As with the previous system only two pathways are possible, a \ce{C-C} or a \ce{C-H} bond breaking, but this time the reactions have different multiplicities (one and six, respectively) and dissimilar entropic contributions.
For all simulations we used adaptive intermolecular reactive bond-ordered potential AIREBO \cite{stuart_reactive_2000}; while classical reactive force fields are not necessarily accurate compared to \textit{ab-initio} or density functional theory methods, they provides a consistent and computationally light framework to test our method.
Since the method itself is not dependent in any way on the underlying potential, this choice affects the result but not the validity of the test.
The theoretical rates were computed by using Eq. \ref{E.Prob}; the values of $\Delta G_i^{\ddagger}$ were obtained by interpolating the \textsc{fe} at the required temperatures; transmission rates were considered equivalent for all the reactions.
For the fast exploration method, we did not apply a thermostat as its efficiency is not constant for all the frequencies and therefore its use can radically influences the results.
The control on the final temperature was obtained by changing the biasing parameters, while still maintaining each added bias relatively small (normally less than 1/10th of $k_BT$ for both Gaussian height and width).
See supplementary materials for more details).
\begin{figure}
\includegraphics[width=3.4in]{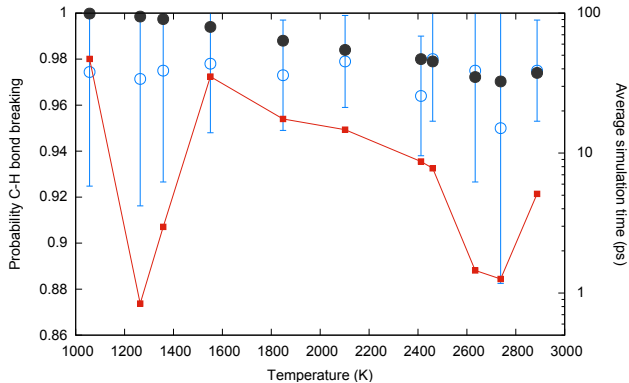}
\caption{Comparison between theoretical (full circles) and computed (empty circles) probability for the \ce{C-H} bond breaking; vertical bars show the 95\% confidence interval (\textit{left y-axis}).
Squares indicate the average simulation time for each system; each point represents the average of 40 to 200 simulations (\textit{right y-axis}).
	\label{F.ethres}}
\end{figure}
As can be seen in Fig. \ref{F.ethres} the predicted probability of a pathway is recovered in a wide range of temperatures, with a minimal computational cost.
Even without any specific optimization, the average simulation time is most of the time 10 ps or less.
Considering a few hundred simulations for each system, the total required simulation time is on the order of a few nanoseconds and can be further  controlled by modifying the number of simulations.

\begin{figure}
\includegraphics[width=3.4in]{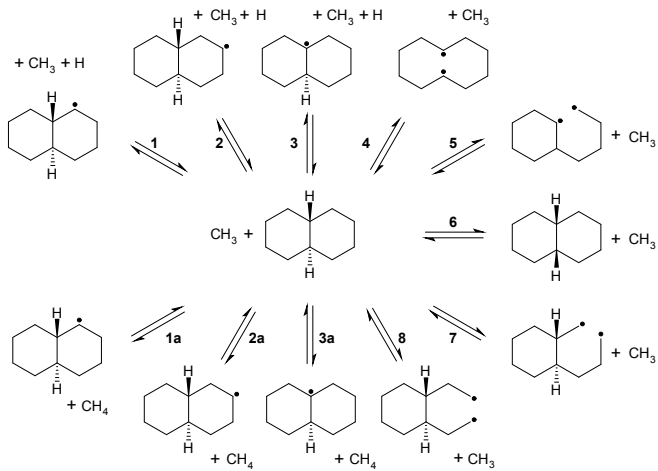}
\caption{Possible reactions of the \textit{t}-decalin + \ce{CH3} system.
Six pathways involve the unassisted (\textbf{1-3}) and assisted
(\textbf{1a-3a}) \ce{C-H} bond breaking, four (\textbf{4,5,7,8}) the \ce{C-C}
cleavage, and one (\textbf{6}) the isomerization to \textit{c}-decalin.
	\label{F.decpaths}}
\end{figure}

\begin{figure}
\includegraphics[width=3.4in]{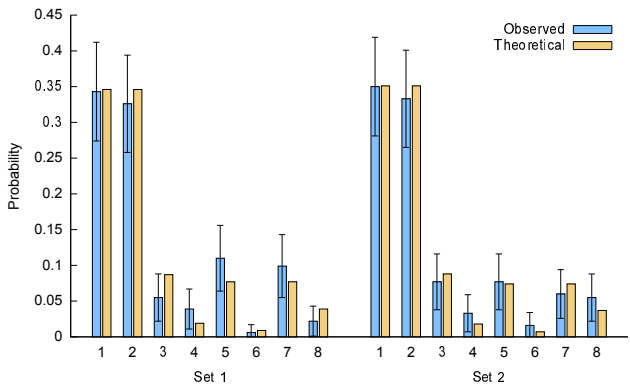}
\caption{Comparison between the observed and theoretical probability of
reactions for the \textit{t}-decalin + \ce{CH3} system for two different set of
conditions.
Reactions are labeled according to Fig. \ref{F.decpaths}.
In both cases the hydrogen abstractions (reactions 1a to 3a) are not reported
due their negligible probability.
Vertical bars represent the 95\% confidence intervals.
	\label{F.decres}}
\end{figure}

As a final application of our method we considered the reactivity of a \textit{t}-decalin molecule in presence of a methyl radical.
The former is used in surrogate fuels and the latter is a common species in combustion environments and plays an important role in H abstraction reactions.
This system can evolve in eleven different products as shown in Fig. \ref{F.decpaths}.
We employed the same general approach used for the ethane study, with the addition of a soft constraint to avoid an excessive increase of the distance between the decalin and the methyl radical.
The results show an excellent agreement for all the pathways (see Fig. \ref{F.decres}) and as before we observe a tremendous speedup, with average simulation times in a range of 2 to 30 ps depending on the biasing parameters.
Notably, our simulations show the complete absence of hydrogen abstraction reactions (1a, 2a, 3a in Fig. \ref{F.decpaths}), which are indicate in the literature as the dominating pathways in such conditions \cite{chae_thermal_2007}.
However, by inspecting the free energy landscape as a function of the distance
of the hydrogen between the carbon of the methyl radical and of the decalin (see supplementary material), one can observe that the abstraction reactions must overcome a barrier of at least 50 kcal/mol more than that needed for the unassisted \ce{C-H} bond breaking and are therefore correctly not observed in a sample of a few hundred runs.
This result is likely related to the choice of a particular version of the AIREBO force field, which does not change the charge distribution as a function of the distance of the methyl radical.
A more recent version of this force field \cite{knippenberg_bond-order_2012} or different reactive potentials may give a more accurate picture of this system's reactivity.
Nonetheless, the validity of our approach is not affected by the choice of the potential.

These tests, in particular the decalin system with eleven different pathways, which include both unimolecular and bimolecular reactions, shows the full potential of our approach.
Within the quite general validity of Eq. \ref{E.TST} our method identifies the subset of the most likely reaction pathways with a minimal computational effort and without neither assumptions on the reactions or transition states nor the need to define different collective variables for each reaction.
The reproducibility of the results independently from the bias and simulation parameters, as well as the generality of the potential energy-SPRINT combination, makes this method suitable for its effective iterative application to complex reaction networks.
Moreover, the options of varying the number of runs for given reactants and blocking specific reactions allow a very efficient exploration even of simple
or partially known reaction networks.

As mentioned above, our acceleration-detection approach can be extended to different classes of systems, with minimal adjustments.
For example, while the potential energy and SPRINT coordinates combination is a general streamlined choice for systems in the gas phase, for other type of reactive networks, like chemical reactions in solutions, the relevant relaxation times, \textit{e.g.}, water reorientation, should be also considered, so that the behavior of the accelerated system is not biased by a specific initial configuration.

\begin{acknowledgments}
This research is funded by the U.S. Air Force Office of Scientific Research (Grant FA9550-13-1-0031) under the technical supervision of Dr. Chiping Li.
The authors thank Michele Ceriotti and Massimiliano Bonomi for the fruitful discussions and valuable suggestions.
\end{acknowledgments}

\bibliography{paper}

\end{document}